\documentclass[12pt]{article}
\pdfoutput=1
\usepackage{Max}

\usepackage{etoolbox}

\allowdisplaybreaks[3]
\title{A comment on 4d and 5d BPS states}
\author{Shamit Kachru}
\author{Max Zimet}
\affil{Stanford Institute for Theoretical Physics,

Stanford University, Stanford, CA 94305 USA}

\date{}

\begin{document}

\maketitle

\begin{abstract}
We discuss a phenomenon in supersymmetric field theory and string theory whereby compactifying one of the dimensions of spacetime on an arbitrarily large circle can cause BPS states to become unstable. We exemplify this by considering 5d $\N=1$ theories on a circle and their embeddings into M-theory via geometric engineering. This implicates a subtle relationship between the BPS states of M-theory on a Calabi-Yau threefold, $X$, and those of type IIA on $X$ with an arbitrary value of the coupling constant. Intuition for this phenomenon is provided by considering F-theory on a complex K3 surface in a limit where it degenerates to a real K3 surface.
\end{abstract}

\newpage
\tableofcontents
\hypersetup{linkcolor=blue}

\section{Introduction}

The spectrum of BPS states -- states whose mass is governed by a central extension of the supersymmetry algebra, minimizing the mass for a given set of charges -- plays an important role in understanding the quantum dynamics of supersymmetric field theories and string theories in diverse dimensions.  One problem of recurring interest,
in relation to subjects ranging from geometric engineering and solutions of field theories with 8 supercharges
to computations of the entropy of extremal black holes, has been the analysis of the BPS spectrum of
M-theory or string theory on a Calabi-Yau threefold $X$.  Several elegant results connect the BPS spectrum of M-theory on $X \times S^1$ to that of type IIA string theory on $X$ (see for instance \cite{4d5d, cardoso:4d5d, DVV, vafa:mWalls}).  
There is, perhaps, an intuition that a BPS state in M-theory on $X$ will descend simply to a BPS state in IIA string
theory on $X$.

\bigskip
In this note, we present a simple example that indicates that, for suitable choices of $X$, M-theory on $X$ can
exhibit BPS states that decay upon further compactifying on a circle with an arbitrarily large radius, $R$. This indicates that the precise relationship between the 5d and 4d BPS spectra of
compactifications with 8 supercharges may be rather nontrivial. In the end, we explain that this decay of BPS states can be given an intuitive explanation via F-theory.

\bigskip
Before continuing, let us briefly contrast this phenomenon with some situations which are superficially similar. First, when spacetime has fewer than four non-compact dimensions there are no states (BPS or otherwise) that are charged under gauge symmetries. This is simply because of the strong infrared effects of gauge fields in low dimensions (e.g., the IR-divergent contribution of a gauge field to the self-energy of a charged particle), which have a classical explanation. So, while it is true that all charged BPS states disappear from the spectrum when one compactifies a 4d theory on a large circle, this is quite different from the inherently quantum effect to which we devote our attention in this paper. Secondly, by compactifying a theory with BPS strings on a circle with an arbitrary radius, one obtains BPS particles from strings wrapping the circle. While this observation plays an important role in the consistency of string dualities \cite{w:comments}, it too has a classical explanation.

\section{The setup}

To find a local setup that will allow us to display the
phenomenon of interest, we consider M-theory on a Calabi-Yau $X$ which is developing a local singularity.  For concreteness, we consider a singularity where a divisor $D$ of topology ${\mathbb P}^1 \times {\mathbb P}^1$ is collapsing to a curve via (say) contraction of the first ${\mathbb P}^1$.  

\bigskip
The physics at such singularities has been analyzed by local methods from the perspective of type IIA string theory in \cite{katz:enhance}, and from the perspective of M theory in \cite{w:phaseTrans}.
In the M-theory picture, one can understand the spectrum of light particles at the singularity as follows.  Consider
an M2-brane wrapping the contracting ${\mathbb P}^1$.  It has a moduli space of vacua given by the base curve
\begin{equation}
{\cal M} \sim {\mathbb P}^1~.
\end{equation}
The low-energy theory on the M2-brane is a supersymmetric quantum mechanics with 4 supercharges.  The supersymmetric
ground states are in correspondence with the cohomology of the moduli space ${\cal M}$.  

\bigskip
More precisely, the M2-brane breaks 4 of the 8 supersymmetries.  Under the $SO(4) \sim SU(2)_L \times SU(2)_R$
little group of a massive particle in 5d, the supercharges transform as 
\begin{equation}
2 \times (1/2,0) ~\oplus ~2 \times (0,1/2)~.
\end{equation}
Without loss of generality, we can take the broken supercharges to be those which transform under $SU(2)_R$.  
Acting on a supersymmetric membrane ground state with these supercharges produces a set of 4 fermion zero modes, 
whose quantization gives the quantum numbers 
$$2 \times (0,0) ~\oplus ~(0,1/2)$$
under the little group.  This alone endows the multiplet of ground states with the quantum numbers of a half hypermultiplet of 5d ${\cal N}=1$ supersymmetry.

\bigskip
The remaining $SU(2)_L$ symmetry also has a role to play.  
The moduli space ${\cal M}$ is K\"ahler.  
The four supercharges of the supersymmetric quantum mechanics
on ${\cal M}$ can be equated with
\begin{equation}
\partial, \bar\partial, \partial^*, \bar\partial^*~,
\end{equation}
and $SU(2)_L$ plays the role of the ``Lefschetz SU(2)'' acting on the cohomology of a K\"ahler manifold.
The action of $J_3$ is
$$J_3 |\psi\rangle = \left( (p+q - {\rm dim}_{\mathbb C} {\cal M})/2 \right) |\psi\rangle, ~~|\psi\rangle \in
H^{p,q}({\cal M})~.$$
$J_+$ and $J_-$ are implemented via multiplication by, or contraction with, the K\"ahler form of ${\cal M}$.

\bigskip
For our specific case with ${\cal M} = {\mathbb P}^1$, we see that the constant form and the top form give rise
to states of $J_3 = \pm {1\over 2}$, filling out a $(1/2,0)$ of $SU(2)_L \times SU(2)_R$.  Therefore, the
full set of quantum numbers of the BPS state arising from the wrapped M2-brane is
\begin{equation}
(1/2,0) \otimes [ 2 \times (0,0) ~\oplus~ (0,1/2)] ~=~2 \times (1/2,0) \oplus (1/2,1/2)~.
\end{equation}
These are the quantum numbers of the states filling out a vector multiplet of 5d ${\cal N}=1$ supersymmetry.

\bigskip
This state, together with the conjugate state obtained by quantization of the anti M2-brane wrapped on this
${\mathbb P}^1$ and the abelian vector arising from reduction of the M-theory three-form on the same cycle,
give rise to a vector multiplet of an enhanced $SU(2)$ gauge symmetry on the locus where the curve collapses.

\bigskip
While the 5d ${\cal N}=1$ supersymmetric pure $SU(2)$ gauge theory doesn't exist as a quantum field theory
(requiring UV completion), M-theory on $X$ provides a suitable regulator. 

\section{Reduction to 4d and quantum dynamics}

Now, consider the physics of $X \times S^1$ in the neighborhood (in moduli space) of the same degeneration.
If the analysis went through as above, one would expect to find 4d ${\cal N}=2$ supersymmetric pure $SU(2)$
gauge theory. 

\bigskip
However, in the quantum theory, there is no point in the moduli space of the 4d gauge theory where $SU(2)$
is restored \cite{sw:theory1}.  This was analyzed from the viewpoint of string theory in \cite{kachru:heteroticK3T2,kachru:geomEng}.
From the gauge theory perspective, given a UV gauge coupling $g_0$, dimensional transmutation gives rise to a
dynamical scale 
\begin{equation}
\Lambda_{SU(2)} \sim M_{UV} ~e^{-{1\over g_0^2}}~.
\end{equation}
  The BPS W-bosons are lifted by quantum
effects before one
reaches a distance of order $\Lambda$ from the would-be origin of the moduli space.
From the string theory perspective, these quantum gauge theory effects are classical -- arising from worldsheet instantons in type IIA string theory, or M2-branes wrapping the base ${\mathbb P}^1$ and the $S^1$ in the M-theory
picture.  

\bigskip
Because of the relationship between 4d and 5d gauge couplings 
\begin{equation}
{1\over g_4^2} \sim {R \over g_5^2}~,
\end{equation}
the W-bosons disappear in a neighborhood of the origin of size $\sim e^{-{R\over g_5^2}}$.  
In particular, for any arbitrarily large $R$, there is a locus in the 5d moduli space close enough to the 
singularity that the $W$-bosons will decay in compactification on $X \times S^1_{R}$.  


\bigskip
From a macroscopic point of view, it is perhaps not surprising that the BPS spectrum can jump on a circle of 
arbitrarily large radius.  The supersymmetric quantum mechanics relevant for analyzing 5d BPS states enjoys an
$SU(2) \times SU(2)$ symmetry.  The supersymmetric quantum mechanics relevant in the 4d problem has reduced symmetry (arising from an $SO(3)$ little group).  The difference in the basic structure of the symmetries of the moduli space one quantizes in searching for BPS 
bound states suggests that the 5d and 4d BPS spectra arising
in M-theory and type IIA on $X$ may enjoy a rather complicated relationship.  

\bigskip
There is a beautiful description of BPS spectra of 4d ${\cal N}=2$ theories in terms of ground states of an auxiliary quiver quantum
mechanics (see for instance \cite{vafa:quivers} for a self-contained discussion).  This can be thought of as the quantum mechanics on wrapped D2-branes in the present example.  It would be very interesting to understand the deformation of the relevant quantum problem from the regime described in \cite{vafa:quivers} to the M-theory limit, to see if a precise relationship between the 4d and 5d systems can be elucidated.

\section{Quantum corrections from F-theory}

In previous sections, we have seen that the BPS W-boson will decay upon compactification from 5d to 4d. However, the phenomenon may still
seem slightly un-intuitive.  A nice feature of string theory is that there is often a duality frame that allows one to intuit quantum corrections, and indeed in this section we will find that that is the case here as well. Our starting point will be type I string
theory compactified on a circle, with a D5-brane wrapping the circle. We consider the world-volume dynamics on the D5-brane. (Duality with the heterotic 5-brane \cite{w:smallInst} now provides a little string theory UV completion, in addition to the gravitational heterotic or type I completions.) T-dualizing the circle yields a type IIA orientifold on $S^1/Z_2$ that is often referred to as type I$'$ theory, where the D5-brane has become a D4-brane. 

\bigskip
In the type I$'$ description, there are 2 O8-planes at the ends of the interval, and 16 D8-branes with variable locations.
When the D4-brane approaches an O8-plane, it classically enjoys an $SU(2)$ gauge symmetry, and as discussed above this persists at the quantum level \cite{s:5d}. Indeed, quantum effects are straightforwardly accounted for using the supergravity description of this setup, and in particular the moduli-dependent coupling constant coincides with the spatially varying string coupling \cite{w:evidence}. However, the field theory on the D4-brane sees only a non-compact region encompassing an end of the interval, as opposed to the entire interval.

\bigskip
We now compactify the D4-brane on a large circle. In order to geometrize the moduli space, as in the 5d theory, we T-dualize this circle, so that we have a type IIB orientifold on a small -- i.e., effectively one-dimensional -- $T^2/Z_2\cong S^2$. The D4-brane has become a D3-brane, so it is not obvious that we still essentially have a 5d field theory; the large fifth dimension emerges from light winding modes, which are present due to the fact that the compactification manifold is small. Nevertheless, this frame allows us to visualize what happens to the moduli space as our field theory becomes four-dimensional. Since BPS states in the field theory correspond to geodesics \cite{sen:BPS} and webs thereof \cite{bergman:FWebs,sethi:FWebs} on the D-brane moduli space, we will find a simple explanation for their decays.

\bigskip
The D3-brane probing an O7-plane was studied in \cite{sen:FOrientifolds,s:fBranes}, where it was explained that quantum corrections are accounted for by F-theory. Specifically, one regards the $S^2$ moduli space of the D3-brane as the $\PP^1$ base of an elliptically fibered K3 surface and the moduli-dependent coupling constant as the complex structure modulus of a fiber. The complex structure of the fibers can be specified by\footnote{More precisely, we can obtain a torus from the following cubic in $\PP^2$:
\be w y^2 = x^3 + f w^2 x + g w^3 \ . \label{eq:torus} \ee
When $w\not=0$, we can scale it away. But, $w=0$ is needed to compactify the torus. Similarly, homogenizing \eqref{eq:fibration} defines a K3 surface via a polynomial of degree 12 in $\mathbb{WP}^3_{1,4,6,1}(w,x,y,z)$. The complex structure of the torus \eqref{eq:torus} is given by
\be j(\tau) = \frac{1}{2} \frac{(24 f)^3}{27g^2 + 4f^3} \ ,\ee
where $j(\tau) = \frac{32 \, \parens{\theta_2^8(\tau) + \theta_3^8(\tau) + \theta_4^8(\tau)}^3}{\theta_2^8(\tau)\theta_3^8(\tau)\theta_4^8(\tau)} = \frac{1}{q} + 744 + 196884q+\ldots$ is the $j$-function, and $q=e^{2\pi i \tau}$.}
\be y^2 = x^3 + f(z) x+ g(z) = (x-e_1(z))(x-e_2(z))(x-e_3(z)) \label{eq:fibration} \ ,\ee
where $f$ and $g$ are polynomials in the coordinate $z$ that parametrizes the $\PP^1$ base with respective degrees 8 and 12. In terms of the 4d $SU(2)$ gauge theory, we have
\be z = \frac{(2\pi)^2}{2}\Tr \phi^2 = (2\pi)^2 w^2 \ ,\ee
where $\phi$ is the scalar in the vector multiplet with eigenvalues $\pm w$; the atypical factor of $(2\pi)^2$ is introduced for later convenience. There are 24 singular fibers, where the discriminant\footnote{This should not be conflated with $\tilde\Delta(\tau) = \eta^{24}(\tau)$, where $\eta(\tau) = q^{1/24} \prod_{n=1}^\infty (1-q^n)$ is the Dedekind $\eta$ function. However, it is the case that if the periods of the fiber are $\omega_1,\omega_2$ with $\tau=\omega_2/\omega_1$, then they are related via $\Delta = -256\parens{\frac{2\pi}{\omega_1}}^{12}\tilde\Delta$, and we also have $f=-240\omega_1^{-4}G_4$ and $g=-2240\omega_1^{-6}G_6$, where $G_k=\sum_{(m,n)\not=(0,0)} \frac{1}{(m+n\tau)^k}$ are the Eisenstein series.}
\be \Delta = 4f^3 + 27g^2 = (e_1-e_2)^2(e_2-e_3)^2(e_3-e_1)^2 \label{eq:discriminant} \ee
vanishes. These correspond to 7-branes.  16 of them can be traced back to the $SO(32)$ gauge symmetry of our type I starting point.
The 8 others arise because the type IIB $T^2/Z_2$ orientifold naturally comes equipped with four O7-planes. Due quantum effects,
each of these splits into two 7-branes (unless it is coincident with 4 D7-branes as well); these two `quantum' 7-branes correspond to the two singularities near the origin in pure $SU(2)$ Seiberg-Witten theory, when viewed by a D3-brane probing the O7 \cite{s:fBranes}.

\bigskip
We now want to take the $S^2$ to be small, which means that the K3 surface will degenerate. To explain the precise degeneration \cite{vafa:realK3}, we find the set of K3 surfaces, $S$, whose second homology lattice splits as
\be H_2(S,\ZZ) \to \Gamma^{17,1}\oplus \Gamma^{1,1}\oplus \Gamma^{1,1}  \label{eq:decompose} \ ,\ee
in the same way as the heterotic string charge lattice on $(S^1)^3$ splits when there are no Wilson lines along two of the circles. In the heterotic frame, the moduli associated to the last two factors in \eqref{eq:decompose} are the radii of these circles, which we wish to take to infinity. We are then left with the familiar moduli space
\be \brackets{ O(\Gamma^{17,1})\backslash O(17,1) / O(17)\times O(1) } \times \RR^+ \label{eq:narain} \ee
of heterotic string theory on $S^1$ (where the final factor is the coupling constant). Similarly, one identifies the homology classes of the K3 surface associated to the final factors in \eqref{eq:decompose}, and this determines the appropriate degeneration. For comparison, if we were to instead study heterotic on $T^2\times S^1$ and take only the radius of the $S^1$ to infinity, we would study K3 surfaces with the decomposition
\be H_2(S,\ZZ) \to \Gamma^{18,2}\oplus\Gamma^{1,1} \label{eq:algebraic} \ ,\ee
namely elliptically fibered K3 surfaces, and take the volume of the fibers to vanish. In this way, we recover the $\PP^1$ moduli space of the D3-brane described above as a limit of the geometry of an M-theory compactification (since M-theory on K3 is dual to heterotic string
theory on $T^3$).

\bigskip
Surfaces satisfying \eqref{eq:decompose} have an elegant classification \cite{vafa:realK3}. First, they are elliptically fibered, as in \eqref{eq:fibration}, where $f$ and $g$ have \emph{real} coefficients. This means that they have a $Z_2$ involution given by complex conjugation of $x,y,$ and $z$.\footnote{Similarly, elliptically fibered K3 surfaces have the involution $y\to -y$.} We refer to the fixed locus (of real dimension 2) of this involution, where $x,y,$ and $z$ are real, as an associated real K3 surface. The degeneration we seek restricts us to such a surface \cite{vafa:F2,vafa:realK3}. Topologically, a real K3 surface is a generally disconnected manifold comprised of a number of spheres plus one Riemann surface \cite{nikulin:realK3}. The case of interest for a dual description of heterotic string theory on $S^1$ has one sphere and a genus 10 surface, and this constrains the functions $f$ and $g$. We refer to this sphere as the real sphere and to the $\PP^1$ base as the $z$-sphere; we also refer to its equator, given by $z=z^*$, as the $z$-line. The $z$-line borders a hole in the genus 10 surface.

\bigskip
Of course, we could switch the $\Gamma^{1,1}$ factors in \eqref{eq:decompose}, which implies that our K3 surface is elliptically fibered in two different ways and both moduli we wish to take to zero correspond to the volumes of the fibers. However, since a hyper-K\"ahler manifold such as K3 has a whole $\PP^1$ of complex structures, these elliptic fibrations need not employ the same one. In fact, with
suitable choices, the fibers of one fibration are calibrated by the holomorphic 2-form $\frac{dz dx}{y}$ -- i.e., they are special Lagrangian. The base of the special Lagrangian fibration is the real sphere, while the base of the elliptic fibration is the familiar base from F-theory.


\bigskip
We now collapse the fibers of the two fibrations.
This collapses both the real sphere and the $z$-sphere to an interval, $S^1/Z_2$. Furthermore, four of the 24 7-branes gravitate to each of the endpoints of this interval, so that the bulk has 16 7-branes. This all is to be expected, since the moduli space in this description should agree with that of the type I$'$ duality frame. In particular, the O8-planes associated to the sets of four 7-branes at the endpoints do not split.

\bigskip
At this point, we can explain qualitatively how the W-boson will be eliminated from the spectrum of a ($S^1$-compactified) D4-brane probing the ``end of the world" O8-plane. In the 5d theory there is a point in moduli space with enhanced gauge symmetry, which in the present picture is where the D3-brane sits on top of a group of 7-branes at an endpoint. The BPS W-boson in this picture is a string stretched from the D3-brane to the 7-branes. However, as soon as the radius, $R$, of the fifth dimension of the field theory becomes finite, these 7-branes will escape from each other along the special Lagrangian fibers which open up. The D3-brane is now alone in the middle of the 7-branes, and so the W-boson disappears from the spectrum. We can also intuit the existence of a wall of marginal stability in the moduli space: if the D3-brane moves far enough away from the 7-branes, then they will appear clumped up again, and so the W-boson will come back into existence (although it will be massive, so the $SU(2)$ gauge symmetry will never be restored). This can all be understood in terms of the existence of certain string webs as a function of one's position in moduli space \cite{fayyazuddin:noGeo,bergman:FWebs,sethi:FWebs}; physically, these strings which comprise BPS states are $(p,q)$-strings of type IIB string theory which are stretched along the $z$-sphere.

\bigskip
We would now like to see this quantitatively. The Seiberg-Witten curve describing a 5d $\N=1$ pure $SU(2)$ gauge theory on a circle of radius $R$ was deduced in \cite{nekrasov:5d,nekrasov:5dInst}. We first define the gauge-invariant modulus
\be U = \frac{1}{2R^2} \Tr P\exp \oint \phi = \frac{1}{R^2} \cosh (2\pi R w) \ ,\ee
where $\phi = A_4 + i\varphi$ combines the component of the gauge field along the circle with the adjoint scalar of the 5d theory. The 5d curve is then the same as the 4d one \cite{sw:theory1},\footnote{The tildes in \eqref{eq:swCurve} serve as a reminder that this equation is not in the form of \eqref{eq:fibration}, i.e., Weierstrass form.}
\be \tilde y^2 = (\tilde x-\Lambda^2)(\tilde x+\Lambda^2)(\tilde x-z) \ , \label{eq:swCurve} \ee
where $\Lambda$ is the dynamically generated scale, if we make the replacement
\be z \to R^2 U^2 - \frac{1}{R^2} \ .\ee
In particular, the singular fibers occur at
\be U = \pm \frac{1}{R^2}\sqrt{1 \pm (R\Lambda)^2} \ .\ee
So, these fibers coincide in the limit $R\to\infty$. Furthermore, away from this limit we find that the number of singular fibers has doubled compared to the 4d theory!\footnote{\cite{nekrasov:5d} argued that this doubling does not occur and that the moduli space is the quotient of the $U$-plane by $U\to -U$, since the theory with only adjoint matter is invariant under the center of the gauge group. However, $U\to -U$ is a \emph{global} symmetry, not a gauge symmetry, since the center of $SU(2)$ acts trivially on $U$. In contrast, if the gauge group were $PSU(2)=SU(2)/Z_2$, then $U$ would be well-defined only up to multiplication by $-1$ (since it is the trace of an element of the complexification of the gauge group).} This is because in the 5d limit, the four corners of $T^2/Z_2$ collapse into the two endpoints of $S^1/Z_2$, and so four 7-branes (i.e., two O7-planes) reside at each endpoint. As we deform away from this limit, the D-brane probe still knows about all four 7-branes. Finally, at $R=0$ these 7-branes are infinitely far apart and the 4d field theory can only see two of the 7-branes. That is, the 4d gauge theory sees only part of the D3-brane's moduli space, similarly to how the 5d field theory sees only the end of the interval. This follows from the fact that taking $R\to 0$ with $z$ finite requires $U = \pm \frac{1}{R^2} + \Oo(R^0)$, and so one cannot interpolate between these two sign choices.

\bigskip
\section*{Acknowledgments}

We thank E. Witten for asking the question which led to this note, and for helpful correspondence.
The research of S.K. was supported in part by a Simons Investigator Award and the National Science Foundation under grant number PHY-1720397.


\providecommand{\href}[2]{#2}\begingroup\raggedright\endgroup

\end{document}